\documentclass[conference]{IEEEtran}
\usepackage{graphicx}
\usepackage{enumitem}
\usepackage{longtable}
\usepackage{array,booktabs}
\usepackage[usenames,dvipsnames,svgnames,table]{xcolor}
\usepackage{url}
\usepackage{adjustbox}
\usepackage{caption}
\usepackage{multirow}

\definecolor{color3}{rgb}{0.45, 0.45, 0.45}
\definecolor{color2}{rgb}{0.55, 0.55, 0.55}
\definecolor{color1}{rgb}{0.72, 0.72, 0.72}
\definecolor{colorNA}{rgb}{0.85, 0.85, 0.85}
\definecolor{colorWorksheet}{rgb}{0.93, 0.57, 0.13}
\definecolor{colorActivity}{rgb}{0.37, 0.62, 0.63}

\begin{document}
	
	\title{An Experimental Study on the Learning Outcome of Teaching Elementary Level Children using Lego Mindstorms EV3 Robotics Education Kit}
	
	\author{\IEEEauthorblockN{Vidushi Chaudhary}
		\IEEEauthorblockA{Samsung, India\\
			vidushi1116@iiitd.ac.in}
		\and
		\IEEEauthorblockN{Vishnu Agrawal}
		\IEEEauthorblockA{EduTECH, India\\
			vishnu.agrawal19@gmail.com}
		\and
		\IEEEauthorblockN{Ashish Sureka}
		\IEEEauthorblockA{ABB, India\\
			ashish.sureka@in.abb.com}}
	\maketitle
	\begin{abstract}
		Skills like computational thinking, problem solving, handling complexity, team-work and project management are essential for future careers and needs to be taught to students at the elementary level itself. Computer programming knowledge and skills, experiencing technology and conducting science and engineering experiments are also important for students at elementary level. However, teaching such skills effectively through active learning can be challenging for educators. In this paper, we present our approach and experiences in teaching such skills to several elementary level children using Lego Mindstorms EV3 robotics education kit. We describe our learning environment consisting of lessons, worksheets, hands-on activities and assessment. We taught students how to design, construct and program robots using components such as motors, sensors, wheels, axles, beams, connectors and gears. Students also gained knowledge on basic programming constructs such as control flow, loops, branches and conditions using a visual programming environment. We carefully observed how students performed various tasks and solved problems. We present experimental results which demonstrates that our teaching methodology consisting of both the course content and pedagogy was effective in imparting the desired skills and knowledge to elementary level children. The students also participated in a competitive World Robot Olympiad India event and qualified during the regional round which is an evidence of the effectiveness of the approach.  
	\end{abstract}
	\IEEEpeerreviewmaketitle
	
	\section{Research Motivation and Aim}
	Robotics construction and programming using robotics education kit for teaching computational thinking, problem solving, programming and engineering skills to elementary school level kids (who fall in the broad range of $4^{th}$ to $7^{th}$ grade and about $8$ to $13$ years old) is a teaching methodology which is gaining popularity in several school curriculum all over the world \cite{karp2010}\cite{kim2006}\cite{petre2004}\cite{garcia2009}\cite{karp2011}\cite{taban2005}\cite{varney2012}. STEM (Science, Technology, Engineering and Mathematics) educators believe that teaching robotics construction and programming is not only effective for teaching robotics, engineering and computational thinking but also fosters essential skills in students like team-work and collaboration (ability to work with others), problem solving, creativity and project management. Lego Mindstorms EV3 is the one of the most popular and widely used robotics education kit in the world. Lego Mindstorms EV3 Education Kit provides several types of parts like the controller (called as brick), motors and sensors (like color, ultrasonic and touch sensors) as well as a visual programming system required for building and programming a variety of robots. Our research motivation is to investigate the application and effectiveness of Lego Mindstorms EV3 for teaching computational thinking, problem solving, programming, team-work and project management to elementary level kids.  Our objective is to examine and observe how students engage in the problem-solving process, how they decompose a larger problem into sub-tasks, how they go about exploring alternate solutions, how they collaborate, communicate and draw conclusions and how they tackle difficulty and complexity. An understanding of such aspects is important to improve and enhance curriculum and teaching methodology.
		\begin{figure*}[!htp]
		\centering
		\includegraphics[scale=0.37, angle=0]{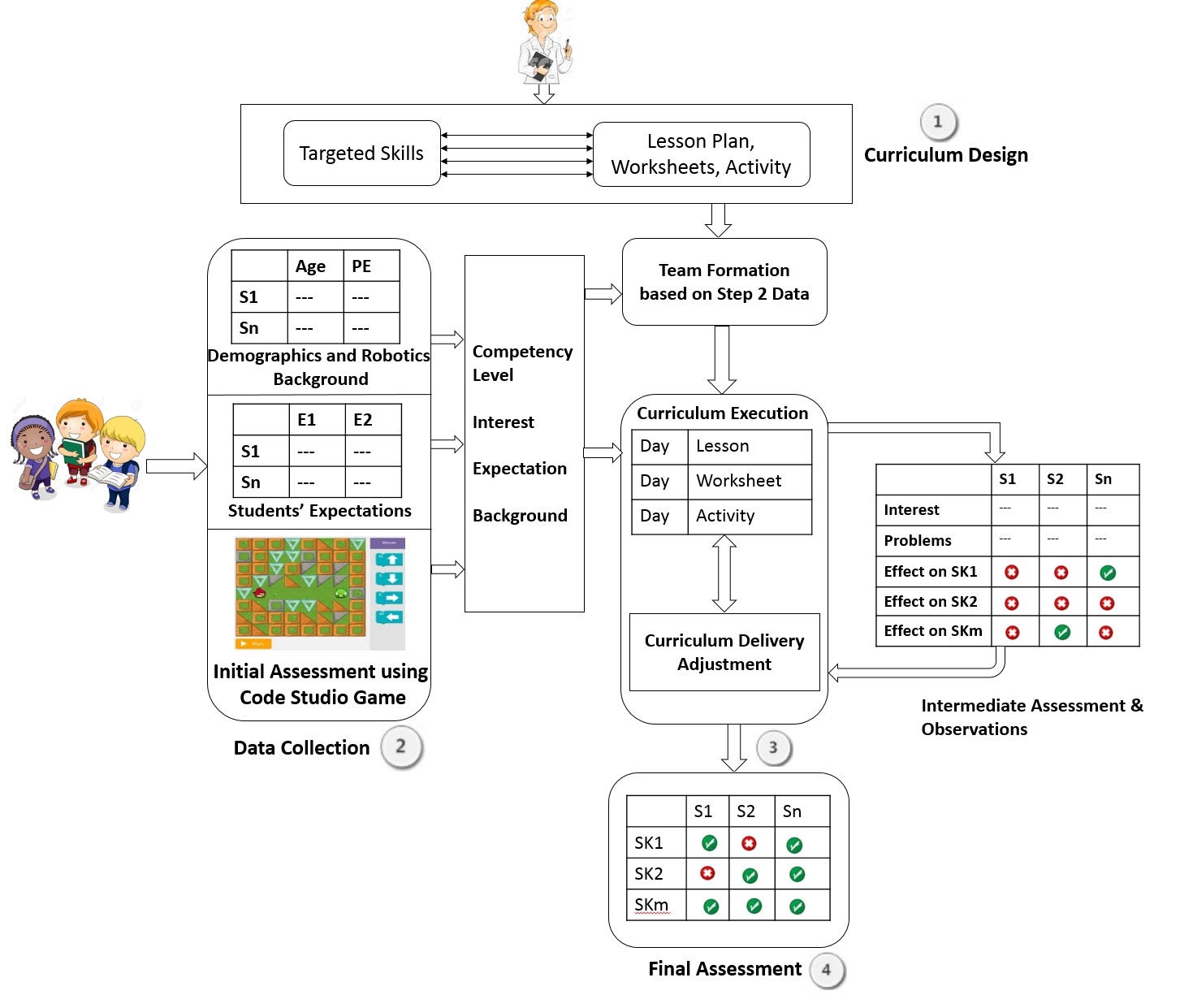} 
		\caption{Research Framework and Methodology}\label{fig:Framework}
	\end{figure*}
	\section{Related Work and Research Contributions}
	In this Section, we present closely related work to our study and present the novel research contributions of our studying context to existing work. Karp et al. present their experiences on the implementation and development of a LEGO robotics engineering outreach program for elementary school students in West Texas \cite{karp2010}. Kim et al. present their approach on educating $C$ language to students using Robotic Invention System $2.0$, a system that helps students to understand the technology of both robot and programming language.\cite{kim2006}. Garcia-Cerezo et al. report on their experience of the $2008$ international summer school on mechatronics based on the LEGO Mindstorms NXT Set \cite{ garcia2009}. Karp et al. describe the evaluation results from an annual LEGO robotics competition for students in elementary and middle schools held at Lubbock, Texas, that aims at increasing interest in science, technology, engineering, and math \cite{karp2011}. 
	
	Taban et al. present a study which describe their experiences on the impacts of basic engineering concepts of LEGO Bricks and Robotics in Coral Academy of Science in Reno, Nevada (a Science, Math and Technology Middle and High School) \cite{taban2005}. Varney et al. describe a program implemented in diverse schools which has been developed for in-school sessions focused around LEGO robotics to foster interest in STEM topics at a young age \cite{varney2012}. Petre et al. conduct observations and interviews with all the participating teams at two LEGO Mindstorms robotics events (one regional, one international) and competitions \cite{petre2004}. Their study reveals that using Lego Mindstorms is effective for teaching programming and engineering to primary and secondary school children in a way that is both well-grounded and generalizable \cite{petre2004}. Galvan et al. share their experiences on the application of Lego Mindstorms kits in the development of teaching curricula for fixed robot manipulators \cite{galvan2006}. In their study, they noticed that the Lego education set is a good tool to analyze robot kinematics and trajectory planning. They present their experiences with a laboratory course designed to address kinematic properties of fixed robots \cite{galvan2006}. 
	\\
	
	\textbf{Research Contributions} The work presented in this paper is an extension of the paper published by the same authors in T4E 2016 \cite{vidushi2016}. Due to a $4$ page limitation in our paper \cite{vidushi2016}, our objective is to provide complete details not covered in the previous paper. To the best of our knowledge, the study presented in this paper is the first experience report and case-study from India on using the robot as a metaphor and using Lego Mindstorms EV3 education kit to teach various skills to elementary school students such as computational thinking, team-work, problem solving and programming. We believe that case studies and experience reports from diverse and heterogeneous countries are important and our work is a contribution to the body of knowledge in application of technology for primary education in developing countries like India.

	While there has been several experience reports on the impact of implementing robotics curriculum as a course in schools, there is a lack of experience reports on summer camps. Summer camps provide a different learning environment and platform than standard courses in schools. Summer camps are of $7$ to $14$ days intense training on a specific topic consisting of students of varying ages but similar interest. Our experimental study is specifically intended to examine the value of robotics summer camp for elementary level children and examine the effectiveness of a short-term course ($7$ to $14$ days) in comparison to a longer duration course. While there has been experience reports on using qualitative methods for evaluating the learning outcome of students in a robotics education program, we use a combination of both quantitative and qualitative methods to verify and evaluate the effectiveness of our program. Our work qualitatively and quantitatively examines the effectiveness of the course curriculum and teaching method designed by us which is customized for our students and first of its kind in terms of the lessons, activities, worksheets, topics, structure and flow of the teaching material. 
	\section{Research Framework and Methodology}
Figure \ref{fig:Framework} shows our research framework and methodology consisting of several steps from curriculum design, surveying students about their expectations, understanding their demographics and prior experience, monitoring their learning, behavior and reactions and finally evaluating the results of the learning process. Each of the steps in Figure \ref{fig:Framework} are covered in detail in the following sections. We create an alignment between the learning environments and the desired skills and knowledge which we want to impart to the students. We create educational indicators and evaluation systems to understand and get feedback about learning outcomes of students and refine the content accordingly as the course progresses. Finally, we acquire data on the overall effectiveness of the summer camp across various parameters. 

\begin{figure*}[!htp]
		\centering
		\includegraphics[scale=0.80, angle=0]{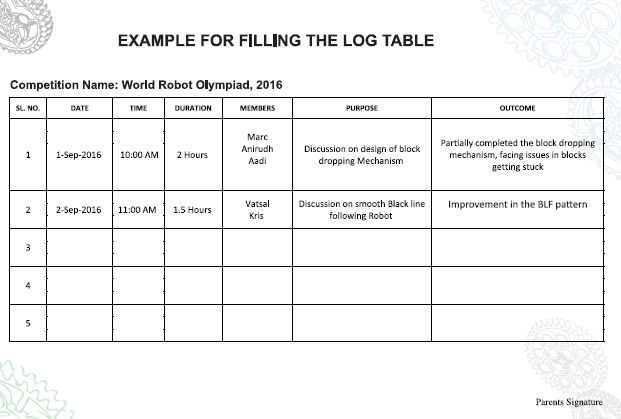}
		\caption{A Log Table used to Record Team Meetings, Date, Time, Duration, Members, Purpose and Outcome}\label{fig:logbook}
\end{figure*}

Figure \ref{fig:logbook} displays a snapshot of the log-book created by us for the purpose of measuring and recording team activities. We emphasized students on the importance of team work and asked to record the data, time, duration, members, purpose and outcome for every team meeting. As shown in Figure \ref{fig:logbook}, we asked students to take weekly signatures from their parents to ensure that students are spending time working in teams and maintaining the log-book in a disciplined manner. Based on student feedback, we learnt that co-operative learning involving students working on their assignments in team is effective in-terms of helping and provide assistance to each-other, sharing credit for the team-work, motivation and fun \cite{slavin1987}. 
	\begin{table*}[ht]
		\renewcommand{\arraystretch}{1.5}
		\caption{Demographics and Prior Robotics Background or Experience of Students [PER: Prior Experience in Robotics]\label{tab:data}}
		\centering
		\begin{tabular}{|c|c|c|c|p{2cm}|p{2cm}|c|p{1.75cm}|c|} \hline
			\centering \textbf{Student} & \centering \textbf{Age} & \centering \textbf{Gender} & \centering \textbf{Grade} & \centering \textbf{Father's Profession} & \centering \textbf{Mother's Profession} & \centering \textbf{PER} & \centering \textbf{Experience Type} & \textbf{Robotics Education Kit} \\ \hline 
			\small \centering{S1} & \small \centering{$11$} & \small \centering{Female} & \small \centering{$6^{th}$} & \small \centering{IT} & \small \centering{IT} & \small \centering{Yes} & \small \centering{Practical} & \small{LEGO Mindstorms EV3} \\ \hline 
			\small \centering{S2} & \small \centering{$12$} & \small \centering{Female} & \small \centering{$7^{th}$} & \small \centering{Non-IT} & \small \centering{Non-IT} & \small \centering{Yes} & \small \centering{Theoretical} & \small{LEGO Mindstorms EV3} \\ \hline 
			\small \centering{S3} & \small \centering{$11$} & \small \centering{Female} & \small \centering{$6^{th}$} & \small \centering{IT} & \small \centering{IT} & \small \centering{Yes} & \small \centering{Practical} & \small{LEGO Mindstorms EV3} \\ \hline 
			\small \centering{S4} & \small \centering{$13$} & \small \centering{Male} & \small \centering{$8^{th}$} & \small \centering{IT} & \small \centering{IT} & \small \centering{Yes} & \small \centering{Theoretical} & \small{LEGO Mindstorms EV3} \\ \hline 
			\small \centering{S5} & \small \centering{$10$} & \small \centering{Male} & \small \centering{$5^{th}$} & \small \centering{IT} & \small \centering{Non-IT} & \small \centering{Yes} & \small \centering{Practical} & \small{LEGO Mindstorms EV3} \\ \hline 
			\small \centering{S6} & \small \centering{$10$} & \small \centering{Male} & \small \centering{$5^{th}$} & \small \centering{IT} & \small \centering{IT} & \small \centering{Yes} & \small \centering{Practical} & \small{LEGO Mindstorms EV3} \\ \hline 
			\small \centering{S7} & \small \centering{$13$} & \small \centering{Male} & \small \centering{$8^{th}$} & \small \centering{IT} & \small \centering{IT} & \small \centering{No} & \small \centering{NA} & \small{NA} \\ \hline 
			\small \centering{S8} & \small \centering{$7$} & \small \centering{Male} & \small \centering{$2^{th}$} & \small \centering{IT} & \small \centering{IT} & \small \centering{No} & \small \centering{NA} & \small{NA} \\ \hline 
			\small \centering{S9} & \small \centering{$13$} & \small \centering{Male} & \small \centering{$8^{th}$} & \small \centering{Non-IT} & \small \centering{Non-IT} & \small \centering{No} & \small \centering{NA} & \small{NA} \\ \hline
		\end{tabular}
	\end{table*}
	\section{Experimental Design and Setup}
	\subsection{Student Demographics and Robotics Background}
	We conduct a summer camp on teaching robotics using Lego Mindstorms EV3 for elementary level children from $8$ May $2016$ to $17$ May $2016$ in Bangalore (India). We created a small batch of $9$ students only so that the instructor can give individual and personalized attention to every student. Diversity in-terms of age, gender and school was one of our goals as we wanted students from diverse backgrounds, cultures and experiences to work together. We first (Step $1$ in Figure \ref{fig:Framework}) collected information about student demographics and robotics background. Table \ref{tab:data} displays information about the demographics and robotics background of the $9$ elementary level students in our summer camp. The youngest student in the batch was $7$ years old and eldest was $13$ years old. There was one $7$ year old student, two students of age $10$, two of age $11$, one of age $12$ and three of age $13$. The average age of the students was $11.12$. The number of male students were twice the number of female students, varying from $2^{nd}$ grade to $8^{th}$ grade in their respective schools. We also captured data about the parent's professions. The reason for doing the same is to discern if the students are already exposed to a science and technology oriented environment at home, as there are higher chances of the children being cognizant of basic computational thinking with such exposures. We observe that approximately $67\%$ of the students have a prior experience on Robotics either as a part of their academic curriculum or because they have previously participated in a robotics summer camp or workshop. We notice that two-thirds of the students who had a prior experience in robotics, had a practical experience, while the rest had a theoretical experience. Also, all the students who had a prior experience used the Lego Mindstorms EV3 Robotics Kit.
\subsection{Student Expectations from Summer Camp}
\begin{figure}[!htp]
		\centering
		\includegraphics[scale=0.40, angle=0]{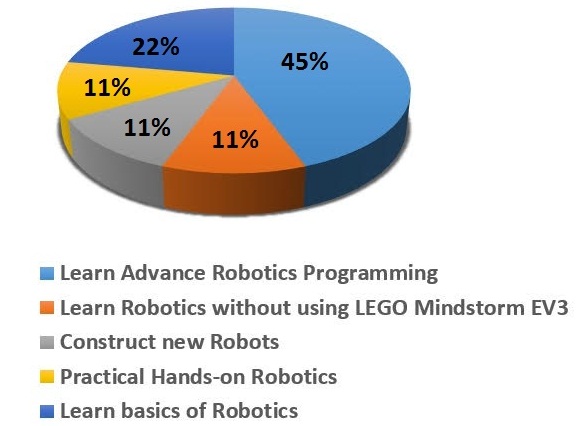}
		\caption{Pie Chart Showing the Expectations of Students from the Course}\label{fig:Expectation}
\end{figure}
Before beginning the summer camp and before monitoring the student learning in the class, we conducted a survey of student expectations from the summer camp. We asked an open question to the student rather than asking a closed question consisting of pre-defined choices.  The data we got from each individual is quite interesting as from $9$ students we received $5$ different categories of expectations. Following are the $5$ categories of expectations defined by the students:
\\
\begin{enumerate}[nolistsep]
\item Learn Advance Robotics Programming
\item Learn Robotics without using LEGO Mindstorms EV3
\item Construct New Robots
\item Practical Hands-on on Robotics
\item Learn Basics of Robotics
\end{enumerate}

The Pie Chart in Figure \ref{fig:Expectation} displays information about the expectations of students from the class. Figure \ref{fig:Expectation} that the most common expectation from students was to learn advance robotics programming (accounts for almost half of the chart). Learning basics of Robotics covers $22\%$ of the pie chart. The lowest percentage was on Learning robotics without using LEGO Mindstorms EV3, Constructing new Robots and Practical Hands-On on Robotics ($11\%$ each). We asked the question on student expectations from the summer camp individually and separately to each student so that they are not influenced by each-other’s responses. We could observe the student fascination for robotics as majority of them said that they want to learn advanced robotics.
\subsection{Competency and Skill Assessment before Camp}
\begin{figure*}
		\centering
		\includegraphics[scale=0.40, angle=0]{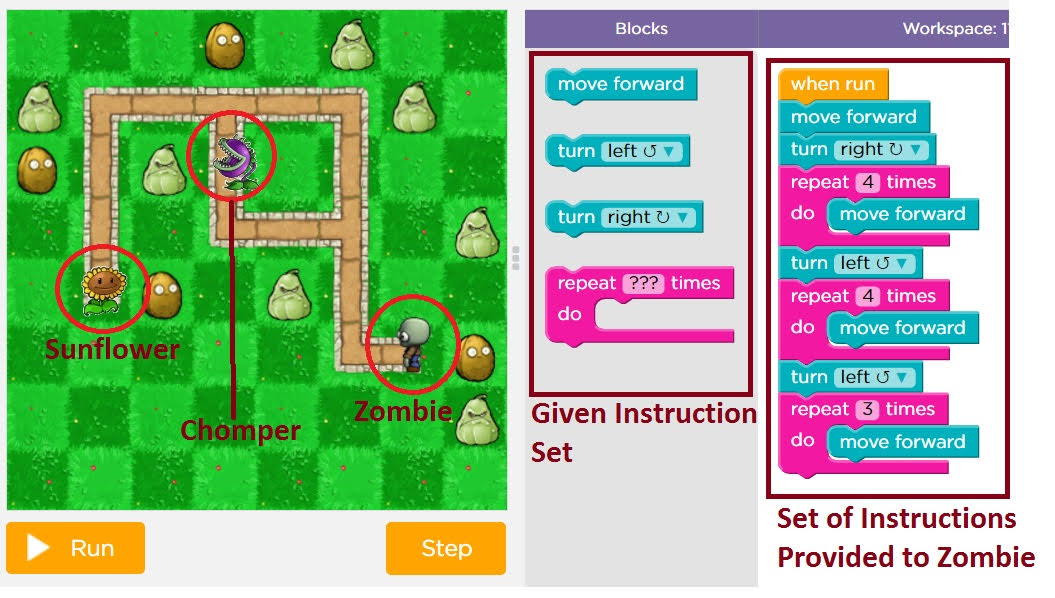} 
		\caption{Game Screenshot}\label{fig:Game_Screenshot}
\end{figure*}
	\begin{figure}[!htp]
		\centering
		\includegraphics[scale=0.40, angle=0]{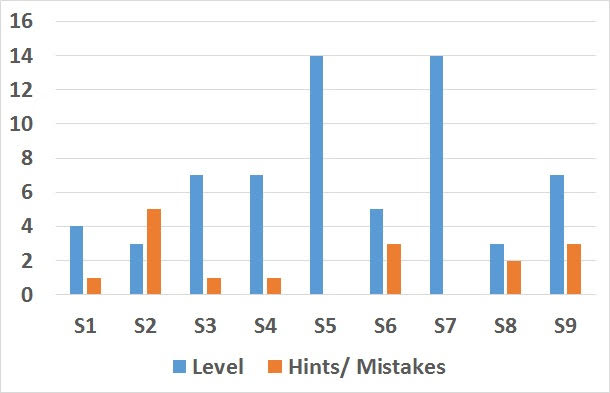} 
		\caption{Bar Chart showing the Assessment Results of Students in-terms of Level Achieved and Number of Hints Provided}\label{fig:Assessment}
	\end{figure}
	Step $2$ (refer to Figure \ref{fig:Framework}) of our research study was to conduct an assessment of the computational and logical thinking skills of the students before the summer camp. We asked students to play an online game which is based on the application of computational thinking. The game comprises of $14$ levels in the increasing level of difficulty. We asked the students to progress one level at a time and qualify as much level as possible. The game is based on the Scratch\footnote{\url{https://scratch.mit.edu}} visual programming language. In the game, there is a zombie, zero or more chompers and a sunflower. The objective of the game is that the zombie should eat all the sunflowers in order to cross the level without coming on the way of chompers. We repetitively instructed the students to make sure that zombie should not come in the path of the chomper as the chompers will eat the zombie. The players are supposed to direct the zombies to the sunflower. The game has $3$ blocks: move forward, turn left and turn right. In addition to these $3$ basic blocks, there is a block of repeat statement. The difficulty level of the game increases with the level number, with more chompers being introduced after a level is crossed. Figure \ref{fig:Game_Screenshot} shows the screenshot of level $8$ of the game.  As shown in Figure \ref{fig:Game_Screenshot}, the game requires students to think logically and create a control flow which requires a set of instructions (such as move, turn and repeat) to be executed. The game used a visual programming language which has similarity to the Lego Mindstorms EV3 programming language. While playing the game, some students were facing difficulties on how to enter instructions. In particular, we observed that some students in understanding the concept of direction and the number of steps required to move to get to the sunflower. We notice that many of them were unable to give $5$ instructions simultaneously or make use of repeat statements. To encourage students to help them qualify the different levels, we provided hints as well. We have hints like providing them a toy demo of giving instructions or if correcting them in diving left or right direction etc. At the end of each student’s game, we collected their assessment data. Out of the $9$ students who played the game, $3$ students could not exceed Level $4$, $2$ students crossed all the levels without any hints and the rest landed up some-where in the range of $5-8$. The bar graph in Figure \ref{fig:Assessment} displays the summary of the assessment result. Our observation and assessment results shows that the computational thinking and programming skills of students required training and improvement which was one the major objectives of the summer camp. 
	\subsection{Course Curriculum, Objectives and Structure}
		\begin{table*}
		\renewcommand{\arraystretch}{1.25}
			\centering
			\caption{Curriculum of the camp-Unit, Lesson Description \& Objective [ CPT- Computational thinking, PRG - Programming, HCD - Handling complexity and Divide Task into Sub-tasks, PRM - project management, TMW - Team work, RBT - Robotics] \label{tab:curriculum}}
			\begin{tabular}{|p{0.5cm}|p{1cm}|p{1.25cm}|p{7cm}|p{0.6cm}p{0.75cm}p{0.75cm}p{0.6cm}p{0.6cm}p{0.85cm}|} \hline
				\centering \textbf {Days} & \centering \textbf {Duration} & \centering \textbf {Unit} & \centering \textbf {Lesson Description \& Objective} & \centering \textbf {CPT} & \centering \textbf {PRG}& \centering \textbf {HCD} & \centering \textbf {PRM} & \centering \textbf {TMW} & \textbf{RBT} \\ \hline 
				\centering \multirow{5}{*}{1} & \centering \small{20} & \centering \small{Lesson 1} & \centering \small{Introduction to robotics, difference between robot, machine \& human} & \cellcolor{colorNA} & \cellcolor{colorNA} & \cellcolor{colorNA}& \cellcolor{colorNA} & \cellcolor{colorNA} & \cellcolor{color1} \\ \cline{2-10}
				& \centering \small{20} & \centering \small{Lesson 2} & \centering \small{Introduction to EV3 hardware (processor, sensors, motors, small \& big parts such as connectors, beams, axles, frames)} & \cellcolor{colorNA} & \cellcolor{colorNA} & \cellcolor{colorNA}& \cellcolor{colorNA} & \cellcolor{colorNA} & \cellcolor{color1} \\ \cline{2-10}
				& \centering \small{30} & \centering \small{Lesson 3} & \centering \small{Introduction to EV3 software (programming blocks such as play, loop switch, move tank, move steering, display} & \cellcolor{colorNA} & \cellcolor{colorNA} & \cellcolor{colorNA}& \cellcolor{colorNA} & \cellcolor{colorNA} & \cellcolor{color1} \\ \cline{2-10}
				& \centering \small{20} & \centering \small \cellcolor{colorWorksheet} {Worksheet} & \centering \small \cellcolor{colorWorksheet} {Questions like: fill in the blanks, match the column \& select correct options among given options. Identifying hardware \& software components of EV3} & \cellcolor{colorNA} & \cellcolor{colorNA} & \cellcolor{colorNA}& \cellcolor{colorNA} & \cellcolor{colorNA} & \cellcolor{color1} \\ \cline{2-10}
				& \centering \small{30} & \centering \small \cellcolor{colorActivity} {Activity} & \centering \small \cellcolor{colorActivity} {Hands-on first time the processor, sensors, motors \& programming} & \cellcolor{colorNA} & \cellcolor{colorNA} & \cellcolor{colorNA}& \cellcolor{colorNA} & \cellcolor{colorNA} & \cellcolor{color1} \\ \cline{1-4}
				\centering \multirow{5}{*}{2} & \centering \small{20} & \centering \small{Lesson 1} & \centering \small{Robot construction using the components} & \cellcolor{colorNA} & \cellcolor{colorNA} & \cellcolor{colorNA}& \cellcolor{colorNA} & \cellcolor{colorNA} & \cellcolor{color1} \\ \cline{2-10}
				& \centering \small{20} & \centering \small{Lesson 2} & \centering \small{Connection of programming blocks with each other \& their usage} & \cellcolor{colorNA} & \cellcolor{colorNA} & \cellcolor{colorNA}& \cellcolor{colorNA} & \cellcolor{colorNA} & \cellcolor{color1} \\ \cline{2-10} 
				& \centering \small{30} & \centering \small{Lesson 3} & \centering \small{Changes in blocks to move the robot in backward direction} & \cellcolor{colorNA} & \cellcolor{colorNA} & \cellcolor{colorNA}& \cellcolor{colorNA} & \cellcolor{colorNA} & \cellcolor{color1} \\ \cline{2-10} 
				& \centering \small{20} & \centering \small \cellcolor{colorWorksheet} {Worksheet} & \centering \small \cellcolor{colorWorksheet} {Drive forward for rotations \& see how much time the robot took to complete the task} & \cellcolor{color1} & \cellcolor{color1} & \cellcolor{color1}& \cellcolor{color1} & \cellcolor{color1} & \cellcolor{color2} \\ \cline{2-10}
				& \centering \small{30} & \centering \small \cellcolor{colorActivity} {Activity} & \centering \small \cellcolor{colorActivity} {Demo program which is already available in the EV3 processor. Front \& back movement using move tank \& move steering blocks by changing the modes (on, off, on for second, on for degree \& on for rotation} & \cellcolor{color1} & \cellcolor{color1} & \cellcolor{color1}& \cellcolor{color1} & \cellcolor{color1} & \cellcolor{color2} \\ \cline{1-4}
				\centering \multirow{5}{*}{3} & \centering \small{20} & \centering \small{Lesson 1} & \centering \small{Concepts of turning (how different vehicles take turns)} & \cellcolor{color1} & \cellcolor{colorNA} & \cellcolor{colorNA}& \cellcolor{colorNA} & \cellcolor{colorNA} & \cellcolor{color2} \\ \cline{2-10}
				& \centering \small{20} & \centering \small{Lesson 2} & \centering \small{Concept of how power of 2 different wheels differ from each other, when any vehicle take turn} & \cellcolor{color1} & \cellcolor{colorNA} & \cellcolor{colorNA}& \cellcolor{colorNA} & \cellcolor{colorNA} & \cellcolor{color2} \\ \cline{2-10}
				& \centering \small{30} & \centering \small{Lesson 3} & \centering \small{Concept of taking left \& right turns by varying the powers of the motors, back left turn, back right turn} & \cellcolor{color1} & \cellcolor{colorNA} & \cellcolor{colorNA}& \cellcolor{colorNA} & \cellcolor{colorNA} & \cellcolor{color2} \\ \cline{2-10} 
				& \centering \small{20} & \centering \small \cellcolor{colorWorksheet} {Worksheet} & \centering \small \cellcolor{colorWorksheet} {Concept of turning the robot right side \& observe the behavior using move tank block} & \cellcolor{color2} & \cellcolor{color2} & \cellcolor{color2}& \cellcolor{color2} & \cellcolor{color2} & \cellcolor{color2} \\ \cline{2-10}
				& \small{30} & \centering \small \cellcolor{colorActivity} {Activity} & \small \cellcolor{colorActivity} {Concept of left turn, right turn, point turn, curve turn, about turn, 45 degree turn, 90 degree turn, 135 degree turn etc.} & \cellcolor{color2} & \cellcolor{color2} & \cellcolor{color2}& \cellcolor{color2} & \cellcolor{color2} & \cellcolor{color2} \\ \cline{1-4}
				\centering \multirow{5}{*}{4} & \centering \small{20} & \centering \small{Lesson 1} & \centering \small{Construction of robotic arm for grabbing the object using medium motor} & \cellcolor{color1} & \cellcolor{colorNA} & \cellcolor{colorNA}& \cellcolor{colorNA} & \cellcolor{colorNA} & \cellcolor{color2} \\ \cline{2-10}
				& \centering \small{20} & \centering \small{Lesson 2} & \centering \small{Medium motor working, how to use the medium motor programming block} & \cellcolor{color1} & \cellcolor{colorNA} & \cellcolor{colorNA}& \cellcolor{colorNA} & \cellcolor{colorNA} & \cellcolor{color2} \\ \cline{2-10}
				& \centering \small{30} & \centering \small{Lesson 3} & \centering \small{Learn about RPM of the motors} & \cellcolor{color1} & \cellcolor{colorNA} & \cellcolor{colorNA}& \cellcolor{colorNA} & \cellcolor{colorNA} & \cellcolor{color2} \\ \cline{2-10}
				& \centering \small{20} & \centering \small \cellcolor{colorWorksheet} {Worksheet} & \centering \small \cellcolor{colorWorksheet} {When you find the object in front of your robot, grab it \& put it aside} & \cellcolor{color2} & \cellcolor{color2} & \cellcolor{color2}& \cellcolor{color2} & \cellcolor{color2} & \cellcolor{color2} \\ \cline{2-10}
				& \centering \small{30} & \centering \small \cellcolor{colorActivity} {Activity} & \centering \small \cellcolor{colorActivity} {Write a program for grabbing the object \& keeping it at the center point} & \cellcolor{color2} & \cellcolor{color2} & \cellcolor{color2}& \cellcolor{color2} & \cellcolor{color2} & \cellcolor{color2} \\ \cline{1-4}
				\centering \multirow{4}{*}{5} & \centering \small{20} & \centering \small{Lesson 1} & \centering \small{Concept of how we can see the things around us, reflection of the light etc.} & \cellcolor{color1} & \cellcolor{colorNA} & \cellcolor{colorNA}& \cellcolor{colorNA} & \cellcolor{colorNA} & \cellcolor{color2} \\ \cline{2-10}
				& \centering \small{20} & \centering \small{Lesson 2} & \centering \small{Working of light sensor} & \cellcolor{color1} & \cellcolor{colorNA} & \cellcolor{colorNA}& \cellcolor{colorNA} & \cellcolor{colorNA} & \cellcolor{color2} \\ \cline{2-10} 
				& \centering \small{20} & \centering \small \cellcolor{colorWorksheet} {Worksheet} & \centering \small \cellcolor{colorWorksheet} {Program robot to stop at the black line} & \cellcolor{color3} & \cellcolor{color3} & \cellcolor{color3}& \cellcolor{color3} & \cellcolor{color3} & \cellcolor{color3} \\ \cline{2-10}
				& \centering \small{60} & \centering \small \cellcolor{colorActivity} {Activity} & \centering \small \cellcolor{colorActivity} {By knowing the RLI (Reflected Light Intensity) of the surface, perform certain tasks} & \cellcolor{color3} & \cellcolor{color3} & \cellcolor{color3} & \cellcolor{color3} & \cellcolor{color3} & \cellcolor{color3} \\ \cline{1-4}
				\centering \multirow{4}{*}{6} & \centering \small{20} & \centering \small{Lesson 1} & \centering \small{Working of GYRO sensor} & \cellcolor{color1} & \cellcolor{colorNA} & \cellcolor{colorNA}& \cellcolor{colorNA} & \cellcolor{colorNA} & \cellcolor{color2} \\ \cline{2-10}
				& \centering \small{20} & \centering \small{Lesson 2} & \centering \small{Understanding the working of GYRO sensor by seeing angle values in port view of EV3 processor} & \cellcolor{color1} & \cellcolor{colorNA} & \cellcolor{colorNA}& \cellcolor{colorNA} & \cellcolor{colorNA} & \cellcolor{color2} \\ \cline{2-10} 
				& \centering \small{20} & \centering \small \cellcolor{colorWorksheet} {Worksheet} & \centering \small \cellcolor{colorWorksheet} {Drive robot for 3 rotations front, take 135 degrees turn \& again drive for 5 rotations} & \cellcolor{color3} & \cellcolor{color3} & \cellcolor{color3} & \cellcolor{color3} & \cellcolor{color3} & \cellcolor{color3} \\ \cline{2-10}
				& \centering \small{60} & \centering \small \cellcolor{colorActivity} {Activity} & \centering \small \cellcolor{colorActivity} {By using GYRO sensors, stop the robot at certain angles, taking exact degrees such as 45, 60, 90, 135, 180 degrees etc.} & \cellcolor{color3} & \cellcolor{color3} & \cellcolor{color3} & \cellcolor{color3} & \cellcolor{color3} & \cellcolor{color3} \\ \hline
			\end{tabular}
		\end{table*}
		\begin{table*}
			\renewcommand{\arraystretch}{1.25}
			\centering
			\begin{tabular}{|p{0.5cm}|p{1cm}|p{1.25cm}|p{7cm}|p{0.6cm}p{0.75cm}p{0.75cm}p{0.6cm}p{0.6cm}p{0.85cm}|} \hline
				\centering \textbf {Days} & \centering \textbf {Duration} & \centering \textbf {Unit} & \centering \textbf {Lesson Description \& Objective} & \centering \textbf {CPT} & \centering \textbf {PRG}& \centering \textbf {HCD} & \centering \textbf {PRM} & \centering \textbf {TMW} & \textbf{RBT} \\ \hline 
				\centering \multirow{5}{*}{7} & \centering \small{20} & \centering \small{Lesson 1} & \centering \small{Concept of ultrasonic waves} & \cellcolor{color1} & \cellcolor{colorNA} & \cellcolor{colorNA}& \cellcolor{colorNA} & \cellcolor{colorNA} & \cellcolor{color2} \\ \cline{2-10}
				& \centering \small{20} & \centering \small{Lesson 2} & \centering \small{What is ultrasonic sensor, Tx \& Rx the ultrasonic sensors} & \cellcolor{color1} & \cellcolor{colorNA} & \cellcolor{colorNA}& \cellcolor{colorNA} & \cellcolor{colorNA} & \cellcolor{color2} \\ \cline{2-10}
				& \centering \small{30} & \centering \small{Lesson 3} & \centering \small{Maximum \& minimum distance which the sensor can detect} & \cellcolor{color1} & \cellcolor{colorNA} & \cellcolor{colorNA}& \cellcolor{colorNA} & \cellcolor{colorNA} & \cellcolor{color2} \\ \cline{2-10}
				& \centering \small{20} & \centering \small \cellcolor{colorWorksheet} {Worksheet} & \centering \small \cellcolor{colorWorksheet} {Program robot in such a way that in starting it will at stop position, as soon as it detects some object in front of it, the robot starts moving} & \cellcolor{color3} & \cellcolor{color3} & \cellcolor{color3} & \cellcolor{color3} & \cellcolor{color3} & \cellcolor{color3} \\ \cline{2-10}
				& \centering \small{30} & \centering \small \cellcolor{colorActivity} {Activity} & \centering \small \cellcolor{colorActivity} {Tasks such as stopping at certain distance, taking turn or any other action when the ultrasonic sensor detects some object at some certain distance} & \cellcolor{color3} & \cellcolor{color3} & \cellcolor{color3} & \cellcolor{color3} & \cellcolor{color3} & \cellcolor{color3} \\ \hline
			\end{tabular}
		\end{table*}
		\begin{table*}[!t]
			\renewcommand{\arraystretch}{1.75}
			\caption{Example of Worksheets Given to Students in Each Class [FIB: Fill in the Blanks, MCQ: Multiple Choice Questions, POP: Programming on Paper]\label{tab:Worksheets}}
			\centering
			\begin{tabular}{|c|c|p{4cm}|c|p{10cm}|} \hline
				\centering \textbf{S.No.} & \centering \textbf{Day} & \centering \textbf{Topic Covered} & \centering \textbf{Type} & {\bfseries \centering Example\par} \\ \hline 
				\small{1} & \small{1} & \small{Identify hardware components} & \small{FIB} & \includegraphics[trim=0 0 0 -5,width=6cm,height=0.8cm]{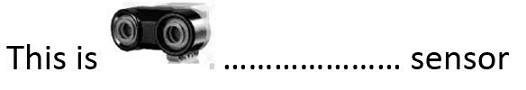} \\[2ex] \hline 
				\small{2} & \small{1} & \small{Identify software components} & \small{MCQ} & \includegraphics[trim=0 0 0 -5,width=6cm,height=0.9cm]{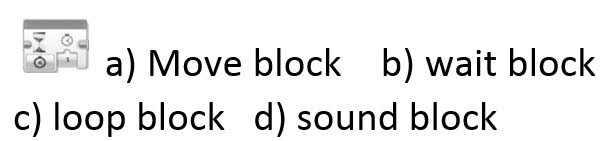} \\[2ex] \hline 
				\small{3} & \small{2} & \small{Motion, movement, steering blocks} & \small{POP} & \small{Drive forward for 2 rotations, using move tank block and observe the time taken by robot to travel} \\ \hline 
				\small{4} & \small{3} & \small{Curve move, turning the robot} & \small{POP} & \small{Drive 2 rotations front then take right turn in the backward direction using move steering block and observe the behavior} \\ \hline 
				\small{5} & \small{4} & \small{Grab the object} & \small{FIB, POP} & \small{How much rotations and degrees the arm should move to grab the object, while driving the robot, grab the object coming in front of it} \\ \hline 
				\small{6} & \small{5} & \small{Program robot to stop at line} & \small{POP} & \small{When robot see the black line, it should take about turn and come back to the same position} \\ \hline 
				\small{7} & \small{6} & \small{Program robot to stop at angle} & \small{POP} & \small{Drive for 5 rotations front, take 90 degrees right turn, drive for 2 rotations forward and then stop the robot} \\ \hline 
				\small{8} & \small{7} & \small{Program robot to stop at object} & \small{FIB, POP} & \small{Write maximum value of the distance an ultrasonic sensor can detect, Program the robot so that it will not move unless and until some object came in front of the ultrasonic sensor distance less than 7cm} \\ \hline 
			\end{tabular}
		\end{table*}
		
We conducted a total of $9$ classes in our summer camp. Each of the $9$ classes was dedicated to a particular topic, with introduction of the basic concepts at the beginning of the class followed by practical hands-on activities. Table \ref{tab:curriculum} displays the course structure, lesson description and mapping of each lesson to the $6$ targeted skills which we wanted to teach to the students. Our objective was to teach $6$ skills to students: CPT- Computational thinking, PRG - Programming, HCD - Handling complexity and Divide Task into Sub-tasks, PRM - project management, TMW - Team work, RBT – Robotics. The shaded cells in Table \ref{tab:curriculum} represents the mapping between lessons and skills. The intensity of the grey color denotes the strength of association between the lesson objective and the targeted skills. As shown in Table \ref{tab:curriculum}, each class had a combination of lecture, practical hands-on activity, tests and work-sheets.
		\begin{figure*}
			\captionsetup{width=.45\textwidth}
			\begin{minipage}[b]{0.50\linewidth}
				\centering
				\includegraphics[scale=0.35, angle=0]{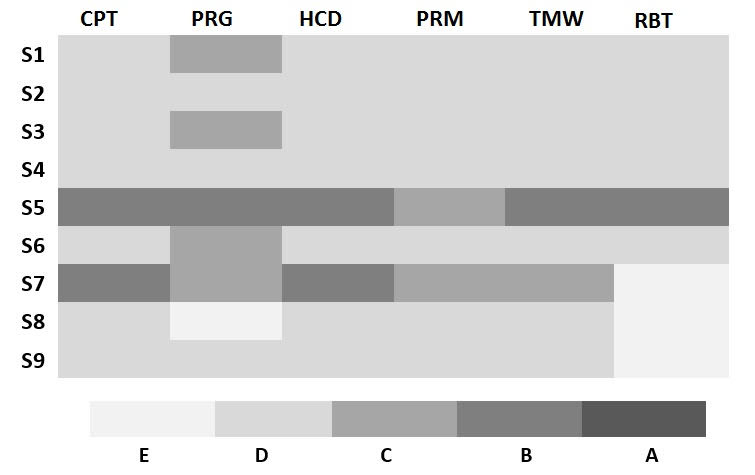}
				\caption{Bertin's Hotel Plot Showing the Grading of Each Student Corresponding to Each Aimed Skill Before the Summer Camp}\label{gradebefore}
			\end{minipage}
			\hspace{-0.50cm} 
			\begin{minipage}[b]{0.50\linewidth}
				\centering
				\includegraphics[scale=0.35, angle=0]{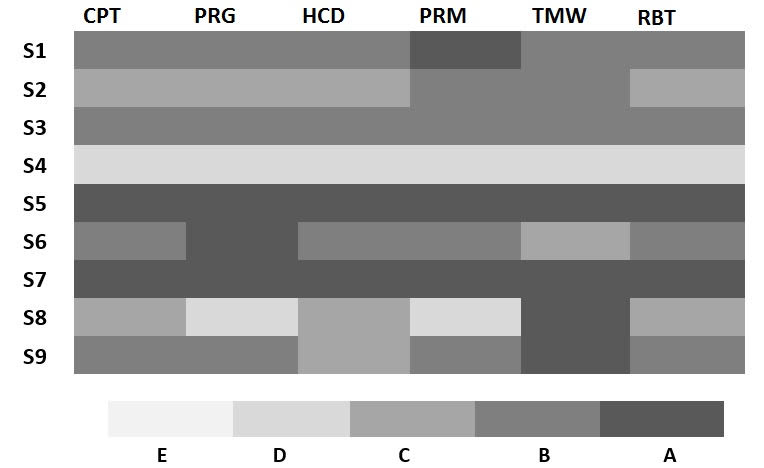}
				\caption{Bertin's Hotel Plot Showing the Grading of Each Student Corresponding to Each Aimed Skill After Completion of the Summer Camp}\label{gradeafter}
			\end{minipage}
		\end{figure*}

In the first class, we introduced students to basics of LEGO Mindstorms EV3 kit, motors, sensors and bricks.  The $2^{nd}$ class introduced students to programming using the LEGO EV3 Mindstorms software. The next class was on robot construction. Students did hands-on on programming and constructed basic models of EV3 robots and also wrote simple programs on how to move a robot. After practicing the basic programming, the objective of the next lesson was to train students on how to command the robot and provide instructions to move in forward and backward direction using tank steering. After completing several robotics programming exercises, the next $5$ classes were dedicated to advanced programming. In $6^{th}$ class, our aim was to teach students how to turn robot using tank steering and different types of curve movements. In $7^{th}$ class, we taught students the construction of the rotor arm to grab and move objects. $8^{th}$ class was to explain them about light sensors and write programs to stop the robot at a line of a given color. In next class, we explained about gyro sensors and using this, we asked students to write a program to stop robot at a certain angle. The last class was on ultrasonic sensors using which students wrote a program to make a robot stop at a certain distance from an object. Table \ref{tab:Worksheets} shows the snapshot of worksheet given to students in each class. Worksheets were given to students to examine their understanding of concepts taught in each class. These worksheets were evaluated by us to know their improvement in robotics field which is one of our targeted skill for the summer camp.
\\\\
\textbf{Novelty Aspects in Content and Procedure}: Our summer camp teaching methodology and course content, structure, activities and worksheets are unique in comparison to previous similar summer camps. We train elementary level students in the complete \textit{engineering design process} (the final capstone project) in which they first imagine and create an architecture of the model, plan, work in teams, divide and break tasks into sub-tasks, construct, program and finally test and improve or refine the model. The model building process also involves free exploration and experimentation (developing divergent thinking) in which the instructor only provides guidance when asked rather than providing direct solutions.
\section{Educational Program Learning Outcome}
		We observe and assess strengths, weaknesses and improvements in student skills through the educational program. The improvement in skills were gradual and the program resulted in encouraging student and learning outcomes. We evaluate the $6$ competencies of all the $9$ students using a $5$ point or letter grade system: $A$ (Excellent), $B$ (Good), $C$ (Average or Acceptable), $D$ (Below Average or Poor) and $E$ (Unacceptable or Fail). The $A$ grade is the highest grade and $E$ grade is the lowest grade. The $9$ students and $6$ competencies constitutes a $9X6$ data or grade matrix. Figure \ref{gradebefore} shows the Bertin plot representing the grades of $9$ students across $6$ competencies. Figure \ref{gradebefore} reveals the grades based on a grey level palette corresponding to the $5$ grades. Figure \ref{gradebefore} reveals that only two students had a $B$ grade on computational thinking and rest everybody had a $D$ grade. The programming skills and knowledge of the students were also below average as they did not have any prior exposure to computer programming. The ability to work effectively in team of $3-4$ members and solving a complex task requiring engagement of more than $30-45$ minutes consisting of designing, constructing and programming robots clearly needed improvement based on our assessment of the students. Issues like difficulty in coming to a consensus or decision, unequal contribution from members, dominance by one member, aggression and negation was visible and the group dynamics certainly needed improvement. 
					\begin{table*}[!t]
			\renewcommand{\arraystretch}{1.75}
			\caption{Students' Feedback [VHF: Very Helpful, NHF: Not Helpful, CPM: Computer Programming, RBC: Robotics Concepts, RCO: Robot Construction] \label{tab:feedback}}
			\centering
			\begin{tabular}{|c|p{0.75cm}|p{0.75cm}|p{0.75cm}|p{0.75cm}|p{0.75cm}|p{0.75cm}|p{0.75cm}|p{0.75cm}|p{1.65cm}|} \hline
				\textbf{Question} & \multicolumn{9}{p{6.75cm}|}{\centering \textbf{Students' Feedback}} \\ \hline
				\textbf{Key Challenges Faced} & \multicolumn{8}{|p{6.75cm}|}{\centering \small{CPM (89\%)}} & \small{RCO (11\%)} \\ \hline
				\textbf{Instructor helpful} & \multicolumn{9}{|p{8.85cm}|}{\centering \small{VHF (100\%)}} \\ \hline
				\textbf{Working in Team} & \multicolumn{7}{|p{5.25cm}|}{\centering \small{VHF (78\%)}} & \multicolumn{2}{|p{3.15cm}|}{\centering \small{NHF (22\%)}} \\ \hline
				\textbf{Takeaways from the Camp} & \multicolumn{4}{|p{3cm}|}{\centering \small{CPM (44.5\%)}} & \multicolumn{4}{|p{3.29cm}|}{\centering \small{RBC (44.5\%)}} & \small{RCO (11\%)} \\ \hline
				\textbf{Would You Again Like to Participate} & \multicolumn{8}{|p{6.75cm}|}{\centering \small{Yes (89\%)}} & \small{No (11\%)} \\ \hline
			\end{tabular}
		\end{table*}
		
		Figure \ref{gradeafter} displays evaluation of the $9$ students across $6$ competencies after the completion of the program. Figure \ref{gradeafter} reveals a substantial improvement in the knowledge and skills of the students. One of the major aims of our educational program was to important team skills and improve the ability of students to get along with their coworkers while working towards a goal. Figure \ref{gradeafter} reveals substantial improvement in collaboration skills of $7$ out of $9$ students. However, we observe that $2$ out of $9$ students were not effective group members and were facing challenges in working in groups. We gauge the team skills of students by observing their interaction with each-other and engagement level towards solving the given task.  Figure \ref{gradeafter} reveals that the computational thinking and programming skills of $6$ out of $9$ students showed substantial improvements. Our assessment shows that $6$ out of $9$ students scored an $A$ or $B$ grade in programming after the completion of the program. We notice that few students learnt how to examine and test if the program is able to perform the given test. They were able to identify and fix errors and were able to write robust programs. Our assessment and experience shows that the experience and exposure received by students through the program improved their computational thinking and programming skills. Our assessment shows that $3$ out of $9$ students scored an $A$ grade on the competency of project and time management. We could clearly notice that some of the students were better in time management and were able to organize themselves better in comparison to other students. Ability to remain focused and not getting distracted and interrupted from the main goal is a skill that required training and we taught students how to stay attentive to their given task. Our result shows an overall improvement in planning, project management and time management skills of the students. 
\\\newline
\noindent \textbf{Robitics Competition Results:} World Robot Olympiad India (WRO India\footnote{\url{http://www.wroindia.org/}}) is a not-for-profit robotics competition held in India since the year 2006 for students between the 9 to 25 years age group. WRO India is a highly competitive, prestigious and one of the largest robotics completion in India. The objective of the event is to encourage science, technology, engineering and mathematics education by giving challenging tasks involving creativity, problem solving and team-work. Teams participating in the competition, design, construct and program robots (using LEGO Robotics Education Kit) which are then evaluated based on the pre-defined scoring system on the day of the competition. WRO India 2016 consists of a regional championship round and a national champion round. The regional competition in Bangalore was on 28 August. Bangalore is one of the top 5 urban agglomeration in India and ranks amongst the world's top few tech-rich cities due to which the competition in Bangalore region is fierce. Our summer camp students formed a team and competed from the Bangalore region consisting of 23 teams at the elementary level. The team qualified for the national championship which is a strong evidence and indicator of the successful learning of the desired skills such as team-work, computational thinking, programming and robotics. 

\section{Student's Course Feedback and Inputs}
We collected student feedback after the completion of the summer camp to evaluate the effectiveness of our approach as well as improve and refine our teaching. We asked students few questions regarding their overall experience in the summer camp. We asked questions like: was working in team helpful, did you find instructor helpful, what are the key challenges faced by you during the camp, what are your key takeaways and learnings from the camp and whether you would like to participate in a follow-up advanced robotics camp. We received encouraging feedback from the students. Table \ref{tab:feedback} shows students feedback for the questions asked. Table \ref{tab:feedback} reveals that $78\%$ of students said working in team is actually helpful to solve the problem effectively while $22\%$ students said it is better to work individually. $100\%$ students said they found the instructor extremely helpful and they learnt a lot during the camp.  While asking about key challenges faced during the summer camp, $89\%$ od students mentioned that they found computer programming quite challenging as it was their first hands-on experience. $11\%$ students felt robot construction was more challenging in comparison to programming. When we asked them about their takeaways from the camp, $56\%$ students said they learnt and experienced computer programming which was new to them, $33\%$ students said they learnt robotics concept very nicely and it was a nice exposure on robotics kit, and $11\%$ students mentioned that they learnt and enjoyed robot construction a lot (assembly and joining of parts). Finally, $89\%$ students said they would like to participate in a follow-up advanced robotics summer camp while $11\%$ students mentioned that they would not like to participate in the advanced robotics camp.
		\section{Conclusion}
		We conclude that using Lego Mindstorms EV3 robotics education kit is effective as a platform and technology to teach and enhance engineering, collaboration, problem solving, time management, computation thinking and problem solving skills for elementary level children. We observe that designing, constructing and programming robots is exciting for students and increases their engagement level. Hands-on assignment and tasks makes the learning both fun and challenging. We conclude that teaching system integration, creative and innovative design from components, parts and connectors is easier and more effective for instructors with a robotics education kit than lecture based approach. Also, computational thinking and programming is easier to teach using Lego Mindstorms EV3 programming system as it is visual drag and drop based rather than text based. Our results shows that elementary level students can understand the concepts of robotics, programming, team-work, modularity, integration and construction from concrete parts or components. We demonstrate a constructivist pedagogy and teaching approach and through our practical experiences support the case of integrating robotics technology into early childhood education at elementary level. We believe that the qualification of students taught in the summer in regional World Robotics Olympiad India championship is an evidence of the effectiveness of the proposed approach and curriculum.
		\bibliographystyle{IEEEtran}  
		\bibliography{T4E2016}  

\begin{thebibliography}{10}
\providecommand{\url}[1]{#1}
\csname url@samestyle\endcsname
\providecommand{\newblock}{\relax}
\providecommand{\bibinfo}[2]{#2}
\providecommand{\BIBentrySTDinterwordspacing}{\spaceskip=0pt\relax}
\providecommand{\BIBentryALTinterwordstretchfactor}{4}
\providecommand{\BIBentryALTinterwordspacing}{\spaceskip=\fontdimen2\font plus
\BIBentryALTinterwordstretchfactor\fontdimen3\font minus
  \fontdimen4\font\relax}
\providecommand{\BIBforeignlanguage}[2]{{%
\expandafter\ifx\csname l@#1\endcsname\relax
\typeout{** WARNING: IEEEtran.bst: No hyphenation pattern has been}%
\typeout{** loaded for the language `#1'. Using the pattern for}%
\typeout{** the default language instead.}%
\else
\language=\csname l@#1\endcsname
\fi
#2}}
\providecommand{\BIBdecl}{\relax}
\BIBdecl

\bibitem{karp2010}
T.~Karp, R.~Gale, L.~A. Lowe, V.~Medina, and E.~Beutlich, ``Generation nxt:
  Building young engineers with legos,'' \emph{IEEE Transactions on Education},
  vol.~53, no.~1, pp. 80--87, 2010.

\bibitem{kim2006}
S.~H. Kim and J.~W. Jeon, ``Educating c language using lego mindstorms robotic
  invention system 2.0,'' in \emph{Proceedings 2006 IEEE International
  Conference on Robotics and Automation, 2006. ICRA 2006.}, 2006, pp. 715--720.

\bibitem{petre2004}
M.~Petre and B.~Price, ``Using robotics to motivate `back door' learning,''
  \emph{Education and Information Technologies}, vol.~9, no.~2, pp. 147--158,
  2004.

\bibitem{garcia2009}
A.~Garc{\'\i}a-Cerezo, J.~G{\'o}mez-de Gabriel, J.~Fern{\'a}ndez-Lozano,
  A.~Mandow, V.~F. Munoz, F.~Vidal-Verd{\'u}, and K.~Janschek, ``Using lego
  robots with labview for a summer school on mechatronics,'' in
  \emph{Mechatronics, 2009. ICM 2009. IEEE International Conference on}.\hskip
  1em plus 0.5em minus 0.4em\relax IEEE, 2009, pp. 1--6.

\bibitem{karp2011}
T.~Karp and A.~Schneider, ``Evaluation of a k-8 lego robotics program,'' in
  \emph{Frontiers in Education Conference (FIE), 2011}.\hskip 1em plus 0.5em
  minus 0.4em\relax IEEE, 2011, pp. T1D--1.

\bibitem{taban2005}
F.~Taban, E.~Acar, I.~Fidan, and A.~Zora, ``Teaching basic engineering concepts
  in a k-12 environment using lego bricks and robotics,'' in \emph{Proceedings
  of the 2005 ASEE Annual Conference and Exposition}, 2005, pp.
  13\,727--13\,736.

\bibitem{varney2012}
M.~W. Varney, A.~Janoudi, D.~M. Aslam, and D.~Graham, ``Building young
  engineers: Tasem for third graders in woodcreek magnet elementary school,''
  \emph{IEEE transactions on education}, vol.~55, no.~1, p.~78, 2012.

\bibitem{galvan2006}
S.~Galvan, D.~Botturi, A.~Castellani, and P.~Fiorini, ``Innovative robotics
  teaching using lego sets,'' in \emph{Proceedings 2006 IEEE International
  Conference on Robotics and Automation, 2006. ICRA 2006.}\hskip 1em plus 0.5em
  minus 0.4em\relax IEEE, 2006, pp. 721--726.

\bibitem{vidushi2016}
V.~Chaudhary, V.~Agarwal, P.~Sureka, and A.~Sureka, ``An experience report on
  teaching programming and computational thinking to elementary level children
  using lego robotics education kits,'' in \emph{International Conference on
  Technology for Education (T4E)}.\hskip 1em plus 0.5em minus 0.4em\relax IEEE,
  2016.

\bibitem{slavin1987}
R.~E. Slavin, ``Cooperative learning and the cooperative school,''
  \emph{DOCUMENT RESUME EA 023 724 Brandt, Ronald S., Ed. Cooperative Learning
  and the Collaborative School: Readings from" Educational Leadership.},
  vol.~45, p.~2, 1987.

\end{thebibliography}
	\end{document}